\documentclass{PoS}

\title{The Roper Puzzle}

\ShortTitle{Roper Puzzle}

\author{\speaker{Keh-Fei Liu}\thanks{A footnote may follow.}\\
        Dept. of Physics and Astronomy, University of Kentucky, Lexington, KY 40506, USA\\
        E-mail: \email{liu@pa.uky.edu}}

\author{Ying Chen\\
        Institute of High Energy Physics, Chinese Academy of Sciences, Beijing 100049, China\\
       E-mail: \email{cheny@ihep.ac.cn}}

\author{Ming Gong, Raza Sufian, and Mingyang Sun\\
       Dept. of Physics and Astronomy, University of Kentucky, Lexington, KY 40506, USA\\ }
 
 \author{Anyi Li\\
            Institute of Nuclear Theory, University of Washington, Seattle, WA 98195, USA\\
        E-mail: \email{anyili0928@gmail.com}}

\abstract{We carried out a calculation of the Roper state with the Sequential Empirical Bayesian (SEB) method with overlap valence fermion on $2+1$-flavor domain-wall fermion configurations on the $24^3 \times 64$ lattice with $a^{-1} = 1.73$ GeV. The light sea quark mass corresponds to a pion mass of 330 MeV. The mass of the Roper, chirally extrapolated to the physical pion mass, is 1404(112) MeV which is consistent with the experimental value at 1440 MeV.
When compared to the Roper state calculation with variational method for Clover and twisted mass
fermions, it is found that the Roper states from SEB with overlap fermion are systematically lower by 400 - 800 MeV for all the quark masses ranging from light to the strange mass region. We study the origin of the difference by exploring the size of the interpolation field in relation to the radial wavefunction of the Roper and also the dynamical influence of the higher Fock space.}

\FullConference{31st International Symposium on Lattice Field Theory - LATTICE 2013\\
		July 29 - August 3, 2013\\
		Mainz, Germany}

\begin{document}

\section{Introduction}   \label{intro}

     The nature of the lowest nucleon excited state, the Roper resonance at 1440
MeV, has been controversial since its discovery. Since the quark model
predictions~\cite{lw83,ci86} of the radial excitation are much higher than the Roper and the fact
that it is very unusual for an S-wave excitation to be even lower than the
P-wave excitation ($S_{11}(1530)$), it has been speculated that the Roper could be a
pentaquark state~\cite{jw04} or a hybrid baryon~\cite{bc83}. However, our earlier
calculation of three-quark interpolation field with overlap fermion on quenched lattices,
which reached the pion mass as low as 200 MeV, revealed that the positive parity excited nucleon is
higher than the P-wave excited state at higher quark mass, but tends to cross below the P-wave state at lower quark mass and can be consistent with experiment at the physical pion
mass~\cite{mcd05}. This is in contrast with other calculations with Wilson 
fermions and chirally-improved fermion which all resulted in positive parity
excited nucleon much higher than the experimental value, mostly above 2
GeV~\cite{lhk07}.  Since the higher Roper masses obtained from the Wilson fermion are based on
variational calculations; whereas, the lower Roper mass calculated with the overlap fermion is based
on the Bayesian approach, the question arises as to whether such a large disparity is 
due to different fitting algorithms to extract the excited state, 
the quenched approximation, or something else. We shall report results on the Roper mass from the
overlap fermion on 2+1-flavor dynamical domain wall fermion configurations with the Bayesian method as well as a variational calculation and compare them with those from the dynamical Clover, chirally improved, and twisted-mass fermions.

\begin{figure}[htb]  \label{Compare_Roper}
  \centering
  {\includegraphics[width=0.7\hsize,height=0.6\textwidth]{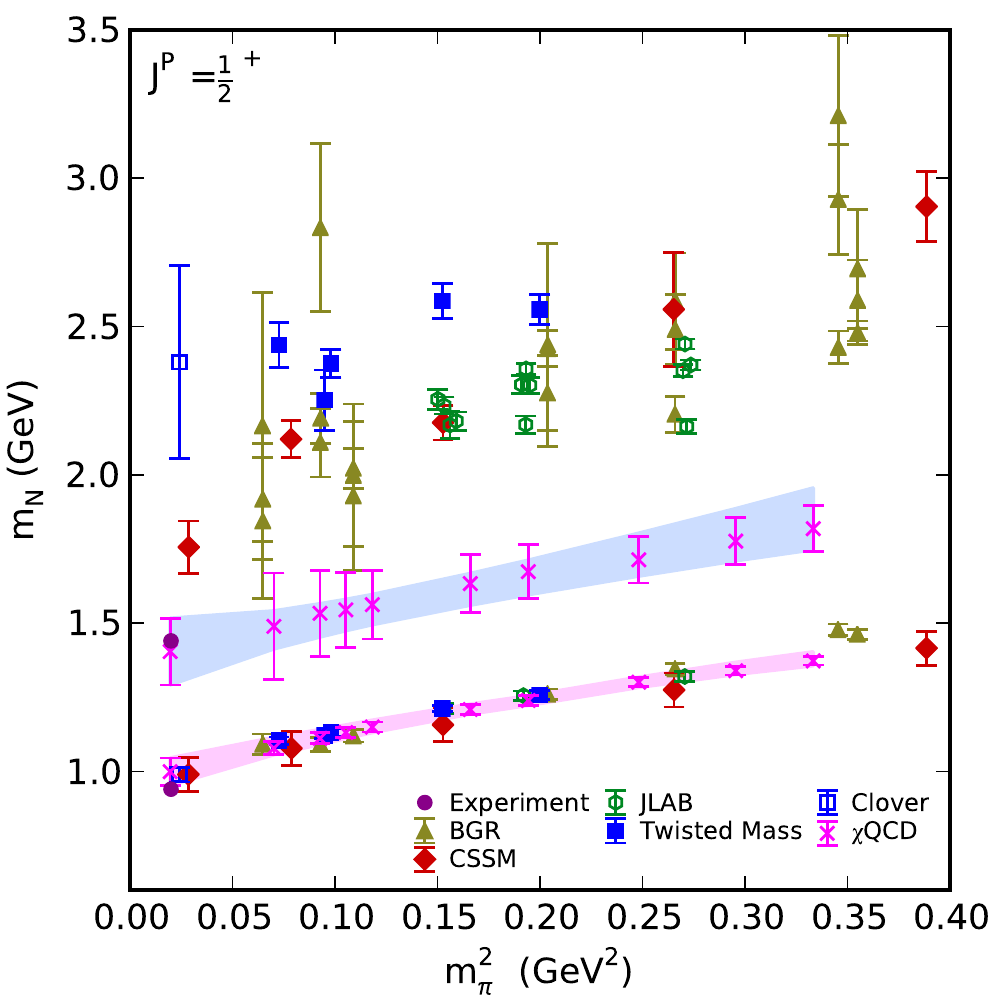}}
  \caption{Nucleon and Roper masses from the present work in comparison with other 
dynamical fermion calculations.}
\end{figure}

   As a first step to reconcile the large disparity in the quenched approximation, we calculate the Roper resonance with the valence overlap fermion on the $2+1$ flavor domain wall fermion configurations on the $24^3 \times 64$ lattice with $a^{-1} = 1.73$ GeV. The light $u/d$ sea quarks correspond to a pion mass of 330 MeV. We have carried out the calculation with the standard 3-quark interpolation field and the Coulomb wall source to reduce the coupling to the P-wave $\pi - N$ scattering state which will require both the pion and nucleon to have one unit of lattice momentum, i.e. $|\vec{p}| = 2\pi/La$ 
for a total zero momentum sink and source. Using the Sequential Empirical Bayes 
Method~\cite{Chen:2004gp},
which has been applied successfully to the excited state calculation previously, such as the Roper
in the quenched approximation~\cite{mcd05}, the $\sigma$ meson~\cite{Mathur:2006bs}, and the pentaquark state study~\cite{Mathur:2004jr},
we obtain the Roper as the first excited state. 
Since we use the mutli-mass algorithm for the overlap inversions, we can obtain propagators for
multiple masses with less than 10\% overhead, we show the results for the nucleon and the Roper for 9 quark masses
in Fig.~\ref{Compare_Roper} as a function of $m_{\pi}^2$. 

For comparison, we have
plotted the nucleon and Roper masses from other calculations with the Clover 
fermions~\cite{edr11,mkl12,akk13}, chirally improved fermions~\cite{elm13}, and twisted mass
fermions~\cite{akk13}. We see that the situation is almost the same as in the quenched approximation, i.e. the nucleon masses agree well in all the calculations, but the first excited nucleon state (i.e. the Roper) from the variational calculations of chirally improved, Clover and twisted-mass fermions are all much higher than those from the overlap fermion with the SEB method. This is true over the range of pion masses below $\sim 600$ MeV.  The only exception is the CSSM result at pion mass of 156 MeV is still $\sim 350$ MeV and 3 sigmas above the experimental value. Our results on the Roper are $\sim 500$ MeV above the nucleon in the pion mass range below 600 MeV. This is consistent with the typical size of the radial excitation of $\sim 500$ MeV for the nucleon, $\Delta$, $\Lambda$, 
and $\Sigma$ as well as heavy quarkoniums. We fit the available data for different quark masses with the form 
$m_{N,R} = m_{N,R}(0) + c_2\, m_{\pi, vv}^2 + c_3\, m_{\pi, vs}^3$, where $m_{\pi, vv}$ is the mass
of the pion made up of a valence quark-antiquark pair with the same quark mass;
while $m_{\pi, vs}$ is the mass of the mixed pion made up of a quark with valence mass and the 
other with the light sea quark mass. At the physical pion limit, we obtain
$m_N = 999(46)$ MeV and $m_R = 1404(112)$ MeV, in good agreement with experiments.

\begin{figure}[h]  
\begin{minipage}{0.5\textwidth}
\centering
  {\includegraphics[width=1.0\hsize,height=0.8\textwidth]{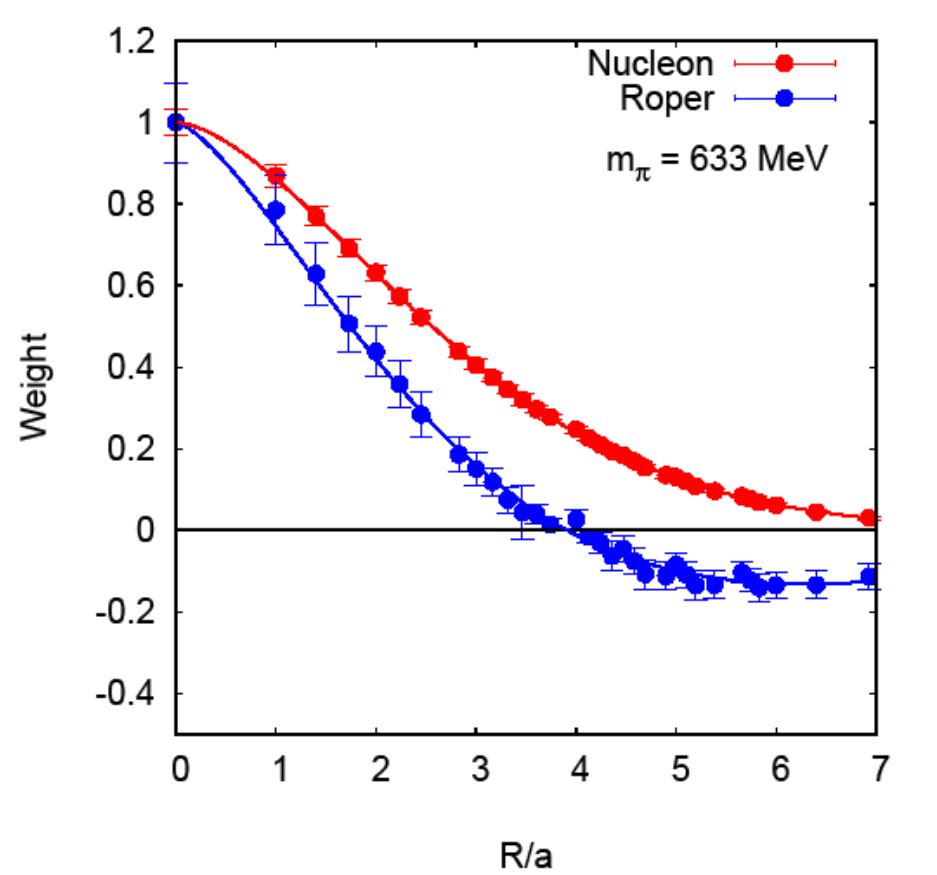}}
\end{minipage}
\begin{minipage}{0.5\textwidth}
\vspace*{-2.2cm}
\centering
  {\includegraphics[width=1.15\hsize,height=1.2\textwidth]{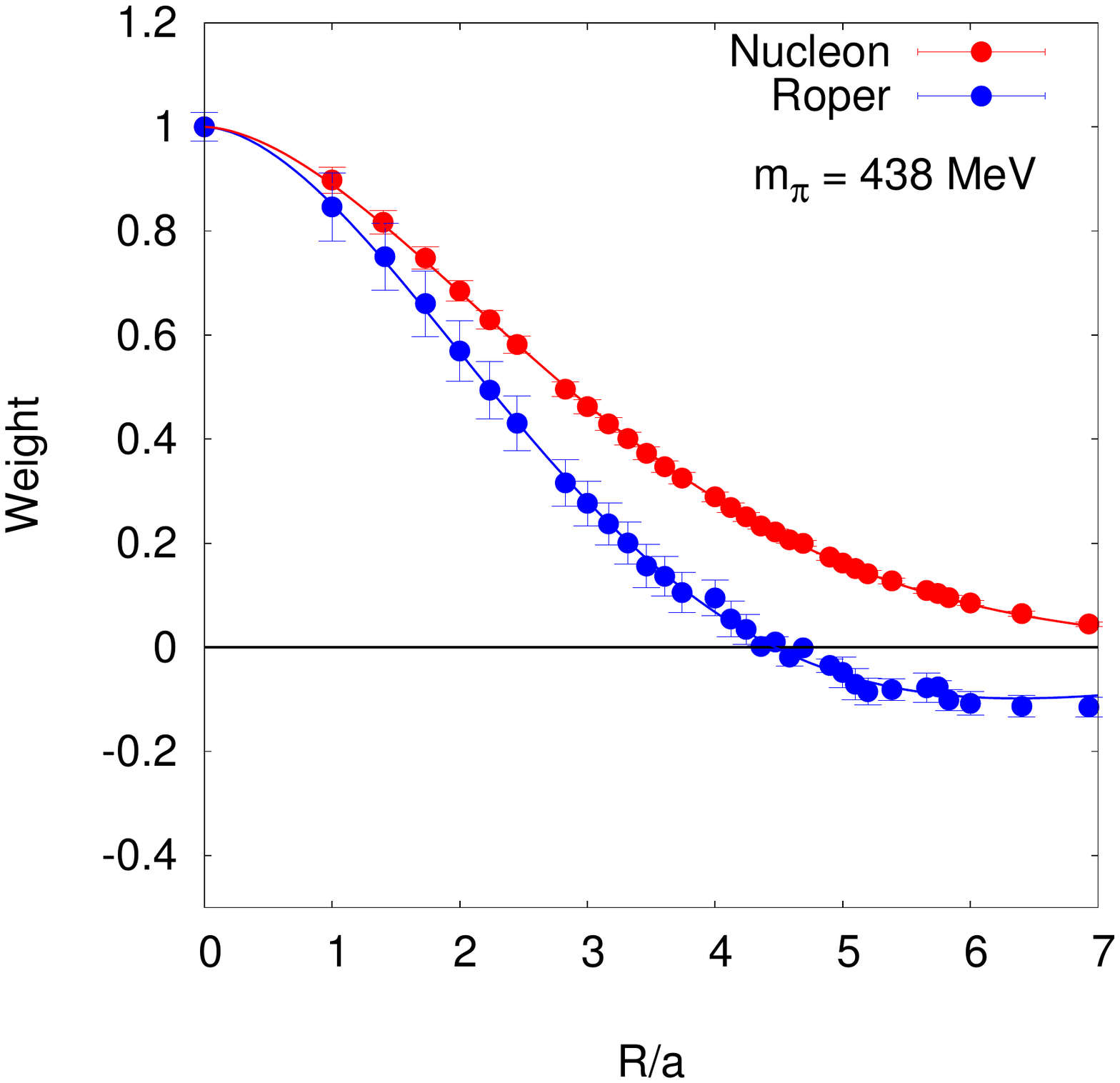}}
 \end{minipage}
  \caption{Nucleon and Roper wavefucntions in the Coulomb gauge for two different pion masses.}
  \label{Roper_WF}
\end{figure}

     As far as the nature of the Roper resonance is concerned, we note that the spectral weight for
Roper from the Coulomb wall source is negative, contrary to that of the nucleon. In addition, we find that the Roper radial wavefunction has a node as shown in Fig.~\ref{Roper_WF} which are obtained by placing two quarks of the sink interpolation field at the same point and the third one at a distance R away on the Coulomb gauge lattice. These are obtained from a quenched calculation with the overlap fermion on a $16^3 \times 28$ lattice with $a = 0.2$ fm~\cite{chen2007}. Since the spectral weight of the Roper with the Coulomb wall source is the product of the matrix element 
$\langle 0|\sum_x \chi (x,x,x)|R\rangle$ for the sink which is positive and the matrix element 
 $\langle R|\sum_{x,y,z} \chi (x,y,z)|0\rangle$ for the wall source which sums up the spatial positions of all three quarks on the wall, this involves an $R^2$ measure which gives more weight to the positions with larger $R$ and results in a negative weight due to the node in Roper as seen in Fig.~\ref{Roper_WF}. Thus, 
both the explicit calculation of the radial wavefunction and the negative spectral weight lend support to the proof that the Roper is primarily a radial excitation of the nucleon with perhaps
some 5-quark $\pi N$ component. The lattice results have ruled out that the Roper is a pure pentaquark state, since such 
a state will have one quark or anti-quark in the P-wave and the rest in the lowest S-wave. This does not lead to a radial node in the wavefunction.

 \begin{figure}[htb]  \label{Variation_small_smear}
\begin{minipage}{0.5\textwidth}
\centering
  {\includegraphics[width=0.9\hsize,height=0.7\textwidth]{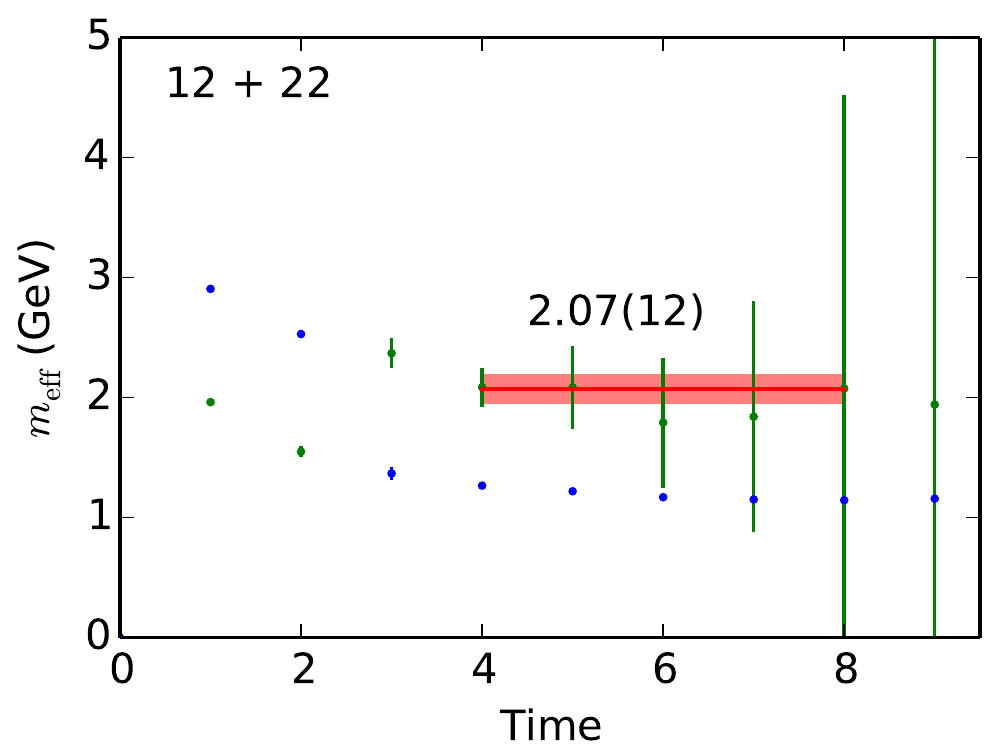}}
\end{minipage}
\begin{minipage}{0.5\textwidth}
\centering
  {\includegraphics[width=0.9\hsize,height=0.7\textwidth]{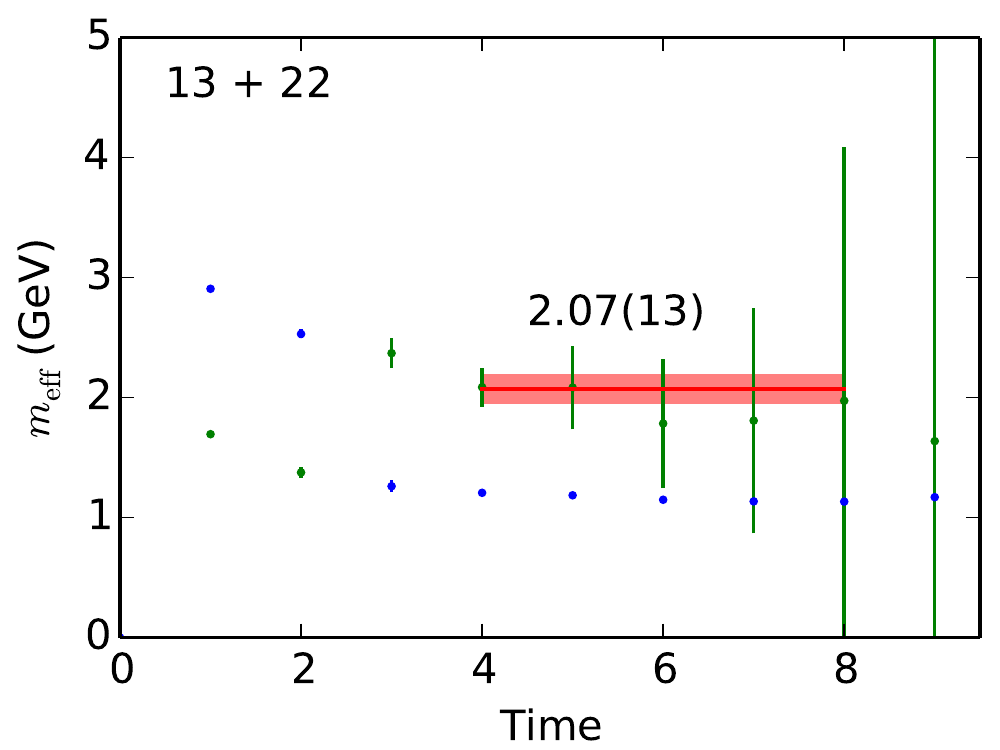}}
 \end{minipage}
  \vspace*{-0.5cm}
  \caption{Nucleon and Roper masses from the variational method with small smeared source.
The $2 \times 2$ variation is characterized by the type of interpolation field (first digit) and the number of smeared quarks (second digit).}
\end{figure}

     Since the node of the Roper occurs at the radial distance of $\sim 0.9$ fm for the light pion 
 ($m_{\pi} = 438$ MeV) case, one would want to have a source whose size is comparable to
 this so that it could differentiate the nucleon from the Roper, provided the radial wavefuntion
 is the major feature which make the two states orthogonal in a variational calculation. Therefore,
 when the size of the interpolation field is smaller than the node distance, we suspect that the first 
 excited state from the variational calculation might have a larger overlap with the 3S state than with the 2S state, which would be the Roper, leading to a higher mass. To test this thesis, we carry out a variational calculation with a mixture of point grid source and a smeared grid source for the three quarks in two different nucleon interpolation fields. This is done on the same $24^3 \times 64$ lattice where the smear size 
 of $r$ = 0.34 fm is substantially smaller than the expected node size of $\sim 0.9$ fm. We have tried
 various combinations with different number of smeared quarks and interpolation fields and find the
 first excited state is  not lower than 2 GeV, about 550 MeV above that from the
 Bayesian fit. This is consistent with our speculation about the size of the interpolation field. We are in the process of repeating the calculation with a smearing size two times larger to check if the SEB result can be verified. If so, it should be a definitive support of the thesis that the mass of the Roper from the variational approach will be higher when the size of the interpolation field is much smaller than that of the radial node of the Roper.

 \begin{figure}[htb]  \label{Var_SEB}
  \centering
  {\includegraphics[width=0.7\hsize,height=0.5\textwidth]{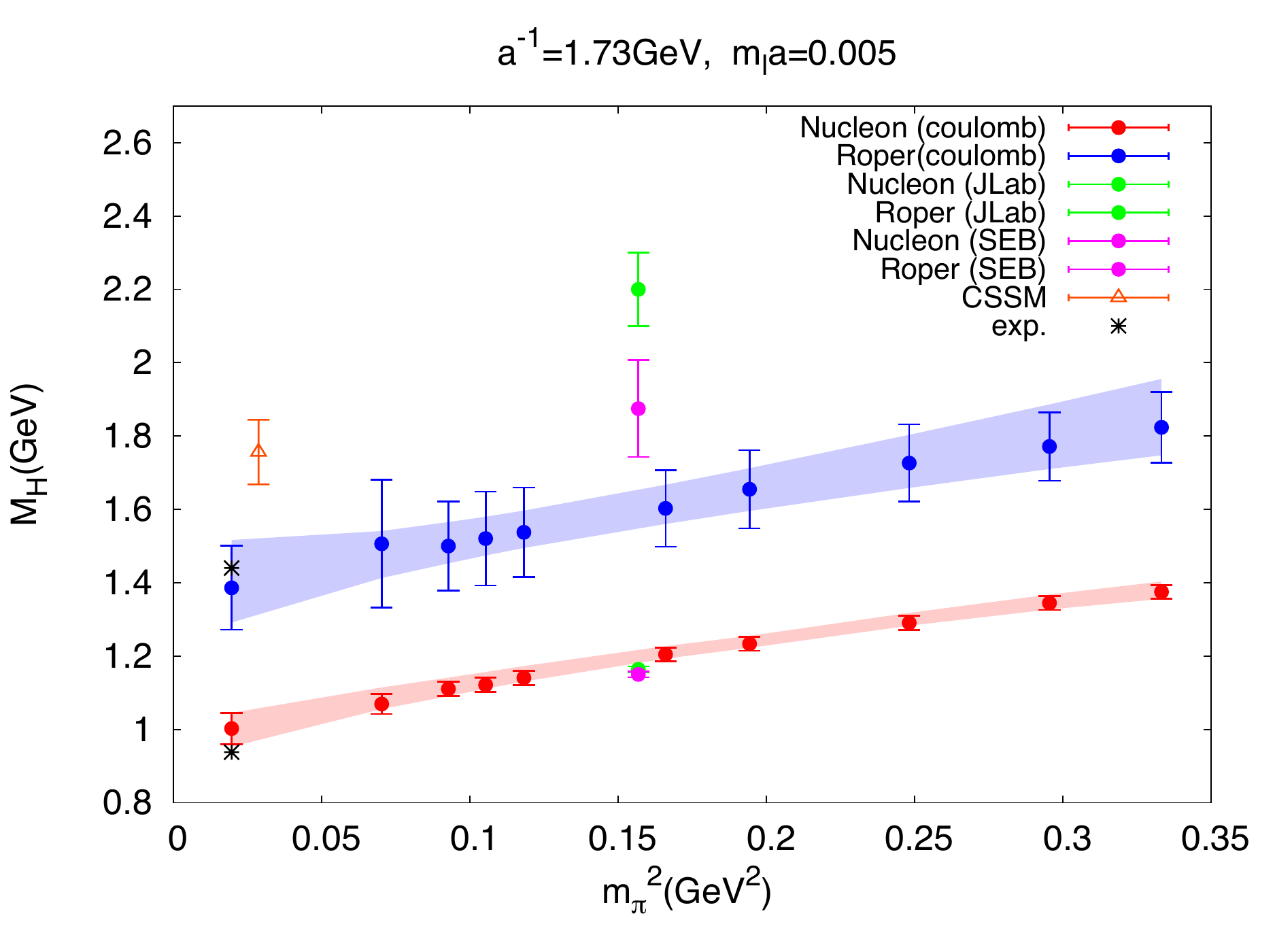}}
\vspace*{-0.5cm}
  \caption{Nucleon and Roper masses. The green points are those from HSC and the purple ones are
from SEB with the same gauge and fermion actions as HSC. The red triangle point is the Roper from CSSM.}
\end{figure}

Next we try to discern how much the difference is due to the algorithm used to determine the Roper state. To this end, we use the gauge configurations that are produced by HSC 
Collaboration~\cite{edr11} to calculate the nucleon 
and the Roper with the SEB method. These are $2+1$ flavor Clover fermion gauge configurations on the anisotropic $24^3 \times 128$ lattice with $a_s = 0.123$fm and the light $u/d$ sea quark corresponds to a pion mass of $\sim 390$ MeV. The SEB fitting of the nucleon and Roper masses are shown in 
Fig.~\ref{Var_SEB} together with the variational results from HSC on the $16^3 \times 128$ lattice with the same lattice spacings and sea quark mass. We see that while the nucleon mass agrees with that from HSC~\cite{edr11}; the Roper, on the other hand, is lower than that from HSC by $\sim 300$ MeV with a $\sim 3 \sigma$ difference. This difference is possibly related to the issue we raised earlier regarding the size of the interpolation field. We note that, with the same SEB approach, there
still appears to be a discrepancy in that the Roper from the Clover fermion is $\sim 350$ MeV above that from the overlap fermion. Similar discrepancy is seen with an extensive variational calculation 
of the Roper state which is carried out at much lighter pion mass of 156 MeV~\cite{mkl12}.  Even at this near physical pion mass, the Roper, calculated at 
$\sim 1.78$ GeV is again about 350 MeV and 3 sigmas above the experimental value. This calculation involves a basis of interpolation fields with varying smearing radius as large as 0.78 fm. Since it is large enough to detect the node of the probability of the Roper wavefunction at 0.77 
fm~\cite{rkl13}, it should not have the issue regarding the size of the interpolation field that we have discussed earlier. 

       To the extent that SEB gives a good fit, the results should be independent of the size of the interpolation field. Therefore, we take the SEB result of the HSC configuration to be the one which would be obtained from the variational approach had a large smeared interpolation field been used. The next question is: why there is a $\sim 350$ MeV discrepancy between the SEB result of the Clover fermion on HSC configurations and that of the overlap fermion at pion mass of 390 MeV and why a similar discrepancy is seen between the Clover fermion result of CSSM and that of the overlap fermion (extrapolated) at close to the physical pion mass. This is still a puzzle. To explore this, we turn to a peculiar feature of the CSSM calculation of the $N(1/2^-)$ state. Since the pion mass is as low as 156 MeV, one would expect the
ground state in the negative-parity channel to be the S-wave $\pi N$ state which is $\sim 450$ MeV
below $S_{11}(1530)$. The surprising result of CSSM is that the lower $\pi N$ is not seen. Instead, the
lowest state is at the mass consistent with $S_{11}(1530)$. This shows that the vacuum to the 5-quark
$\pi N$ matrix element via the 3-quark interpolation field $\chi_{N,3q}$, i.e. 
$\langle \pi N|\chi_{N,3q}|0\rangle$ is small. As a result, the lower $\pi N$ state will emerge at a later time than the fitting window of the calculation. 
In contrast, it is known from the quenched approximation with overlap fermion that the $\eta' N$ ghost state (which appears at the mass of $m_{\pi} + m_N$) becomes prominent for pion mass below 300 MeV~\cite{mcd05}. Our latest calculation of the negative-parity nucleon on the $32^3 \times 64$ domain wall configurations ($a^{-1} = 1.4$ GeV and pion mass at 170 MeV) with the valence overlap fermion clearly reveals the $\pi N$ as the ground state. 


Since the lattice spacings, the sizes of the lattice, and quark masses are similar between the HSC and RBC configurations, it is not easy to fathom why there should be such a sizable discrepancy. We shall
attempt to put forward an explanation and suggest ways to check it. Let's consider a
quantum mechanical system with the hamiltonian $H = H_0 + \lambda H_1$, where $\lambda H_1$  can be considered a perturbation to $H_0$ due to the weak coupling $\lambda$. In second order perturbation, the ground state $E_0$ of $H_0$ will be shifted downward by 
\begin{equation}
\Delta E = - \sum_{n \neq 0}\frac{|\langle n|\lambda H_1|0\rangle|^2}{E_n - E_0},
\end{equation}
independent of the sign of $\lambda$. An extensive model of this kind has been constructed to
study the  $N^*$ resonance in $\pi N$ scattering partial waves~\cite{JuliaDiaz:2009ww}. The $H_0$
is for the bare $N$ and $\Delta$ and $\lambda H_1$ describes the meson-baryon reaction channels including $\pi N, \eta N,$ and $\pi\pi N$ which has $\pi \Delta, \rho N$, and $\sigma N$ resonant components. This model fits the $\pi N$ scattering data well in various channels below 2 GeV. It is found~\cite{Suzuki:2009nj} in the $P_{11}$ channel, the meson-baryon transition is strong which shifts the bare $1/2^+\, N^*$ at 1763 MeV 
down to ${\rm (Re M_R, -Im M_R)}$ = (1357, 76) MeV which corresponds to the $P_{11}$ pole of $N^*(1440)$.
This is a shift of $\sim 400$ MeV in mass due to the meson-baryon coupling. 

In the context of lattice calculation, one can picture $H_0$ as the part describing the 3-quark dynamics 
and the $\lambda H_1$ the coupling between the 3-quark and 5-quark states (e.g. $\pi N, \eta N$, etc). However, the total hamiltonian which includes dynamics for all Fock spaces is prescribed by fermion action. It is possible that different lattice fermions can have different dynamics and, hence, different $\lambda$ at a finite lattice spacing $a$. Thus, the discrepancy we have observed so far may well be an $O(a^2)$ effect. To check this scenario at a given $a$, one could calculate the transition matrix element
\begin{equation}
T_{3q\rightarrow 5q} = \langle 0|\chi_{N, 3q}|\pi N\rangle,
\end{equation}
with the same 3-quark interpolation field for the nucleon, $\chi_{N, 3q}$, 
in both the negative- and positive-parity channels and compare them in different fermion actions to see if there is any sizable difference.

     A related quantity one can also calculate is the overlap integral between the nucleon and Roper wavefunction as discussed earlier,
\begin{equation}
O_{RN} = \int d^3 R \psi^*_R(R) \psi_N(R),
\end{equation}
where $\psi_R$ and $\psi_N$ are normalized Roper and nucleon wavefunctions.
Since they are expected to be orthogonal for heavy quarks, the overlap should be close to
zero. We have indeed verified this with our quenched data~\cite{chen2007}. When the quark mass becomes lighter, we observe that the Roper node moves out and the overlap $O_{RN}$ increases. For example, $O_{RN} = 0.30/0.59 $ for $m_{\pi} = 633/438$ MeV (left /right panel of Fig.~\ref{Roper_WF}). This, we believe, reflects the fact that the 5-quark (e.g. $\pi N, \eta N$) Fock components becomes more important as the quark mass gets lighter. This is consistent with the phenomenological meson-baryon coupled channel model~\cite{Suzuki:2009nj} and the suggestion that it is the large coupling to 5-quark states from the 3-quark interpolation field in the overlap fermion that is responsible for the lower Roper state than those calculated with the Clover fermions. It would be useful to compare the overlap integral $O_{RN}$ with that of the Clover fermion to check this conjecture.

This work is partially supported by U.S. Department of Energy grant DE-FG05-84ER40154.
   We thank Alexandrou for providing us the variational calculation results in Fig.~\ref{Compare_Roper}.

\end{document}